\documentstyle[aps,epsf,preprint,graphicx,prl]{revtex}
\begin{document} 
\draft
\def\beq{\begin{equation}}
\def\eeq{\end{equation}}
\def\beqn{\begin{eqnarray}}
\def\eeqn{\end{eqnarray}}
\def\ed{\end{document}}
\def\x {{\bf x}}

\title{Phase separations of bosonic mixtures in optical lattices\\ from
macroscopic to microscopic scales}

\author{Ofir E. Alon\footnote{E-mail: ofir@tc.pci.uni-heidelberg.de}, 
Alexej I. Streltsov and Lorenz S. Cederbaum}

\address{Theoretische Chemie, Physikalisch-Chemisches Institut, Universit\"at Heidelberg,\\
Im Neuenheimer Feld 229, D-69120 Heidelberg, Germany}

\maketitle
\begin{abstract}
Mixtures of cold bosonic atoms in optical lattices undergo phase separations
on different length scales with increasing inter-species repulsion.
As a general rule, the stronger the intra-species interactions,
the shorter is this length scale. 
The wealth of phenomena is documented by illustrative
examples on both superfluids and Mott-insulators.
\end{abstract}


The first experimental realization of the superfluid to Mott-insulator
phase transition of cold bosonic atoms in optical lattices (OLs) \cite{IB1_nature}, 
following the theoretical suggestion in \cite{Jaksch1_PRL},
has boosted the community to study the physics of bosons in OLs, 
see, e.g., the recent review \cite{Markus_review} and references therein.
Recently, mixtures of bosonic species in OLs have attracted ample attention, see, e.g., 
Refs.~\cite{Boninsegni,Demler_Zhou,Svis13,Paredes_Cirac,Cazalilla_Ho,Altman,Ziegler,Zheng_real,Girvin_BB}.
Obviously, the physics of bosonic mixtures trapped in OLs 
is more intricate than that of the single-species case 
\cite{IB1_nature,Jaksch1_PRL,Markus_review,Fisher_PRB,OL_us}. 
For instance, due to the inter-species interaction additional quantum phases
appear \cite{Demler_Zhou,Svis13,Paredes_Cirac,Altman}.

Phase separation is a basic phenomenon distinguishable particles can undergo with increase mutual repulsion.
How does a mixture of bosonic atoms undergo phase separation in an OL, instantly or, perhaps, step by step?
Is there a difference between phase separation of two superfluids or of two Mott-insulators? 
In answering these questions, we show below that a bosonic mixture in OLs
undergo a wealth of {\it phase separations} on different {\it length scales}.
As a general rule, the stronger the intra-species interactions, the shorter 
the length scale in which phase separation in OLs can take place. 
More details are given below.
Phase separation, by definition, happens in {\it real space},
which is `where' we choose to attack this problem. 

Our starting point is the many-body Hamiltonian describing interacting $N=N_A+N_B$ bosons,
$N_A$ bosons of type $A$ and $N_B$ bosons of type $B$, in an optical lattice,
\beq\label{Ham}
  H_{AB} = H_A + H_B + W_{AB}.
\eeq
Here $H_A$ is the usual single-species Hamiltonian (for the $A$-type bosons),
containing one-body $h_A$ and two-body interaction terms.
Similarity, $H_B$ describes the $B$-type bosons.
$W_{AB}$ contains the inter-species two-body interaction terms. 
In what follows we employ the common contact interaction between all bosons
in the mixture, 
see, e.g., Ref.~\cite{BB_HO_Esry_Bigelow},
where the corresponding interaction strengths are denoted by $g_A$, $g_B$ and $g_{AB}$
and are proportional to the intra- and inter-species s-wave scattering lengths.

As mentioned above, we are to attack the properties of a bosonic mixture in the OL in real space.
Recently, a multi-orbital best-mean-field ansatz for single-species bosonic systems
has been derived \cite{LA_OAL_PLA} and led us to predict 
a wealth of quantum phases and excitations of strongly-interacting bosons in OLs \cite{OL_us} and 
novel phenomena associated with fragmentation and fermionization of bosons in traps \cite{frag_OL_fer}.
Anticipating that an approach similar in spirit would be valuable for mixtures of bosons in OLs,
we proceed to construct one.

Consider the Hamiltonian (\ref{Ham}) in the absence of intra-species interaction, $W_{AB}=0$.
The two species are then independent and can be treated separately. 
The total wavefunction of the system $\Psi_{AB}=\Psi_A\Psi_B$ is just 
a product of the wavefunctions $\Psi_A$ and $\Psi_B$ of the individual species.
Take species $A$ and write a multi-orbital ansatz for $\Psi_A$.
This amounts to attaching an orbital to each of the $N_A$ bosonic atoms.
Generally, we may take $n_1$ $A$-type bosons to reside in one orbital, $\phi_1(\x)$,
$n_2$ $A$-type bosons to reside in a second orbital, $\phi_2(\x)$, and so on,
distributing the $N_A$ atoms among $1 \le n_{orb} \le N_A$ orthonormal orbitals.
More formally, we use the ansatz 
$\Psi_A=\hat {\cal S} \Big\{\phi_1(\x_1) \cdots \phi_1(\x_{n_1})
 \phi_2(\x_{n_1+1}) \cdots \phi_2(\x_{n_1+n_2}) \cdots
 \phi_{n_{orb}}(\x_{N_A})\Big\}$, where 
$\hat{\cal S}$ is the symmetrization operator
and $\x_i$ is the coordinate of the $i$-th $A$-type boson. 
Taking the expectation value of the Hamiltonian $H_{A}$ with respect to $\Psi_{A}$
leads to the single-species multi-orbital energy functional \cite{LA_OAL_PLA}:
\beqn\label{BMF}
E_A= & & \sum_i^{n_{orb}}
\left\{n_i  \int \phi_i^\ast(\x) h_A(\x) \phi_i(\x) d\x
+ g_A \frac{n_i(n_i-1)}{2} \int |\phi_i(\x)|^4 d\x\right\} \nonumber \\
 & & + \sum_{i<j}^{n_{orb}} 2 g_A n_i n_j \int |\phi_i(\x)|^2 |\phi_j(\x)|^2 d\x. \
\eeqn
In order to find the ground state of $A$ bosons 
in the subspace of all possible configurations $\Psi_{A}$,
one has to minimize $E_A$ with respect to its arguments.
These are the number $n_{orb}$ of orbitals in which the $A$-type bosons reside, 
the occupations $\{n_i\}$ of these orbitals
and the orbitals $\{\phi_i\}$ themselves which have to be determined {\it self-consistently}.
Performing this minimization results 
in a system of $n_{orb}$ coupled, non-linear equations
which is solved {\it self-consistently} \cite{OL_us,LA_OAL_PLA}.
Next, the $B$-type bosons are treated in {\it exactly} the same manner,
and one only needs to introduce a separate notation for the $B$-type 
orbitals, their number and occupations, and the $B$ bosons' energy-functional
which we respectively denote by
$\{\psi_p\}$, $m_{orb}$ and $\{m_p\}$, and $E_B$.
We point out that due to the distinguishability of
the $A$- and $B$-type bosons, no relations are assumed between the $\phi$'s
and $\psi$'s orbitals which, therefore, are {\it a priori} allowed to overlap. 

With all ingredients at hand, let us switch on the inter-species interaction $W_{AB}$.
We now employ $\Psi_{AB}=\Psi_A\Psi_B$ as the ansatz for the wavefunction of the mixture {\it including}
inter-species interaction,
analogously to utilizing $\Psi_A$ as the ansatz for interacting single-species bosons \cite{LA_OAL_PLA}.
Taking the expectation value of the full Hamiltonian $H_{AB}$ with respect to $\Psi_{AB}$
readily leads to the multi-orbital Bose-Bose energy functional:
\beq\label{BB_energy_functional}
 E_{AB}= E_A +  E_B 
+\sum_i^{n_{orb}} \sum_p^{m_{orb}} g_{AB}\, n_i m_p \int |\phi_i(\x)|^2 |\psi_p(\x)|^2 d\x, \
\eeq
Finding the minimum of $E_{AB}$
amounts to minimize it with respect to the variational parameters of $E_A$ and of $E_B$ 
which were already discussed above.
But now, the inter-species interaction $W_{AB}$ couples {\it all} orbitals
which, as we shall see below, leads to very many effects.
Before proceeding, it is gratifying to recognize that 
the two-component Gross-Pitaevskii approach, see, e.g., Ref.~\cite{BB_HO_Esry_Bigelow},
is a specific case of $E_{AB}$ ($\Psi_{AB}$) when 
all $A$-type orbitals are alike and 
all $B$-type orbitals are alike, namely $n_{orb}=m_{orb}=1$.

The multi-orbital Bose-Bose energy functional (\ref{BB_energy_functional}) 
can be employed to study demixing in optical lattices
by following the ground state as the inter-species repulsion grows. 
Let us examine some of its properties.
Consider a given bosonic mixture with fixed intra-species interaction 
strengths $g_A$, $g_B$ and inter-species repulsion $g_{AB}$ (we shall only consider repulsive mixtures here).
We can minimize the energy functional (\ref{BB_energy_functional}) and find the ground state, 
i.e., the set of orbitals  $\{\phi_i\}$, $\{\psi_p\}$
and their respective occupations $\{n_i\}$, $\{m_p\}$ that minimize $E_{AB}$.
For $g_{AB} \ll g_A,g_B$ the single-species terms $E_A$ and $E_B$ 
dominate $E_{AB}$ because the inter-species interaction term 
(the third term on the r.h.s. of Eq.~(\ref{BB_energy_functional})) is much smaller.
Consequently, both species can completely mix and spread all over the optical lattice.
Moreover, the physics of each species is determined by its own parameters almost independently 
of the other species state.
Increasing $g_{AB}$ influences the physics of {\it both} species. 
In order to reduce the inter-species interaction energy in (\ref{BB_energy_functional}),
the orbitals $\{\phi_i\}$ and $\{\psi_p\}$ reduce their overlap in space.
When $g_{AB} \gg g_A,g_B$ we expect this overlap to vanish, which means
that the two species completely separate and occupy different regions in space.
In between these two extreme cases, as we shall see below,
there can be many intriguing possibilities of phase separation on smaller length scales.
In fact, even when the two species fully separate in space
there are differences, e.g., between two superfluids and two Mott-insulators.

In a general bosonic mixture,
the $A$ and $B$ bosons have different masses, filling factors and
intra-species interactions, and they can possibly feel different optical potentials.
Here, to capture and explore the richness of possibilities,
still maintaining a coherent line of exposition, 
we consider both $A$- and $B$-type bosons to have the same mass, 
the same intra-species repulsion $g_A,g_B$ 
and filling factors, denoted hereafter by $f_A,f_B$, 
and to feel the same optical lattice. 
Additionally, we examine mixtures in one-dimensional optical lattices.
It is convenient to introduce the dimensionless coordinate $x$
in which the one-body Hamiltonian (of each species) reads 
$h = -\frac{1}{2} \frac{d^2}{dx^2} + \frac{1}{2} E_R \cos^2(x)$,
where $E_R$ stands for the depth of the optical lattice in recoil energy units.
Throughout this work we take $E_R=20$. 
The coupling constants $g_{A}$,$g_{B}$ and $g_{AB}$ 
are now related to the scattering lengths and the confining parameters of the transverse directions 
\cite{Maxim_PRL}. 
In one-dimension one often expresses the 
strength of (intra-particle) interaction 
in terms of the dimensionless parameter $\gamma_{A}$ which is the ratio of interaction and kinetic energies.
In the above units one has 
$\gamma_A=\pi\frac{g_A}{f_A}$ and similarly for the $B$-type atoms.

Let us begin by studying the phase separation in the simplest case, i.e., 
of two (weakly-interacting) superfluids.
Taking $g_{A}=g_{B}=10^{-3}$ and $g_{AB}=10^{-4}$ and, for instance, $f_{A,B}=\frac{1}{2}$ 
we minimize the energy functional (\ref{BB_energy_functional}) 
and find that it is minimized when all $A$ bosons reside in one orbital 
and all $B$ bosons reside in another orbital, 
i.e., $n_{orb}=m_{orb}=1$. 
Therefore, for these parameters
the best mean-field of the bosonic mixtures is the two-component Gross-Pitaevskii 
approach \cite{BB_HO_Esry_Bigelow}. 
How do the corresponding orbitals look like? 
They are delocalized over the whole lattice, see Fig.~1A. 
Obviously, no demixing occurs. 
Next we consider a much stronger repulsion, $g_{AB}=1$, and reminimize $E_{AB}$.
We again find that Eq.~(\ref{BB_energy_functional}) is minimized with one orbital per species;
the orbitals, however, now separate in space, see Fig.~1B.
This is the simplest phase separation possible. 
It is interesting to examine the interface region between the now separated species, see Fig.~1B. 
It shows smaller densities which grow and saturate as one moves away from the center of the interface.
Clearly, by lowering the species' densities in the interface region 
the energy is minimized. 
Why can the density of each species at all be lowered?
Because after phase separation had occurred each species remains a superfluid 
and is, thus, compressible. 
Repeating the above calculations for the same interaction strengths
and the filling factors of $f_{A,B}=1$
we obtain the same scenario described above and in Figs.~1A,B.      

The above phase separation of two superfluids is on the longest possible length scale, i.e., 
of the order of the lattice size.
We notice that the length scale available for phase separation 
is simply determined by the sizes of the orbitals in which the different species reside.
This brings about the following idea. 
If we could make the sizes of these
orbitals {\it smaller} then we might be able to realize states where
the bosonic species are phase-separated on a {\it shorter} length scale.
How can we make the size of the orbitals smaller?
By increasing the intra-species interactions each species would independently fragment, 
namely it would occupy more orbitals of smaller sizes, see Ref.~\cite{frag_OL_fer}.
Along these lines, we consider two strongly-interacting superfluids, specifically 
we take $g_A=g_B=0.1$ with $g_{AB}=0.1$, and $f_{A,B}=\frac{1}{4}$.
Minimizing $E_{AB}$ we find that: 
(1) each  boson resides in a different orbital,
(2) each orbital is spread over two lattice sites, and  
(3) orbitals of different species are alternating in position, see Fig.~2A.
In other words, phase separation on a much shorter length scale has been achieved.
Taking now a much stronger inter-species repulsion, $g_{AB}=10$,
the strongly-interacting superfluids fully demix, see Fig.~2B.
Next, we take the same interaction parameters but smaller filling factors, $f_{A,B}=\frac{1}{6}$,
to allow for more space for each boson.
Indeed, the size of each 'domain'
is now three lattice sites, see Figs.~2C,D.
Additionally, we observe that the spatial densities are maximal in the center of
each 'domain' and minimal in the interface of two 'domains',
in reminiscence of the finding for two
weakly-interacting superfluids shown in Fig.~1B.
Decreasing the densities to $f_{A,B}=\frac{1}{2p}$, where $p$ is an integer,
it is anticipated that larger 'domains' whose sizes are $p$ lattice sites can be formed.  

So far, we considered phase separations on length scales which are larger than the size of one unit cell. 
Let us move towards phase separations on smaller sizes.
Consider the case of $f_{A,B}=\frac{1}{2}$ and take $g_A=g_B=0.1$.
By minimizing $E_{AB}$ for $g_{AB}=0$ we obtained that 
all bosons reside in different orbitals
and that each orbital is spread over two lattice sites.
Of course, 
this is a system of two independent species, each being a strongly-interacting superfluid.
What happens as we switch on $g_{AB}$?
For a small inter-species repulsion, $g_{AB}=10^{-3}$, 
all orbitals are still spread over two lattice sites and arrange in a staggered conformation
as shown in Fig.~1C where
there is half an $A$ and half a $B$ boson per site.
For $g_{AB}=0.1$, all orbitals shrink to the size
of a single site and the system is now in a Mott-insulator state where    
the $A$ and $B$ atoms arrange in alternating sites, see Fig.~1D.
In other words, phase separation on a length scale of one lattice size has happened.
It is relevant to mention that an alternating pattern 
has been obtained by describing the bosonic mixture 
in the OL within an effective-spin Hamiltonian \cite{Altman}.   
Finally, taking a much larger inter-species repulsion, $g_{AB}=10$,
a complete phase separation of the Mott-insulator state of the mixture emerges, see Fig.~1E.
It is instructive to examine the interface region of the two separated species,
which shows no visible lowered density, 
contrary to the case of two phase-separated superfluids shown in Fig.~1B.
The reason is that the two-species Mott-insulator is incompressible.
  
We have also investigated a mixture with the filling factors $f_{A,B}=1$ and $g_A=g_B=0.1$.
By minimizing $E_{AB}$ for $g_{AB}=0$ we find  
that each boson resides in a different orbital
located on a single lattice site.
For $g_{AB}=10^{-3}$ 
we still have two Mott-insulator phases which
sit atop one another, like in Fig.~3A.
For $g_{AB}=0.1$ we obtain that $E_{AB}$ is minimized by
having two bosons per orbital.
Hence, half the number of orbitals is required 
and, side by side, phase separation within neighboring sites occurs similarly to Fig.~1D.
Increasing the inter-species repulsion enforces particles of the same
kind to share space and thereby enlarge the separation length-scale size.
Finally, for a much larger inter-species repulsion, $g_{AB}=10$,
full phase separation occurs like in Fig.~1E, 
but now with two bosons per orbital.

Till now, 
we have described phase separations scenarios on length scales down to a single lattice site.
As it turns, 
we cannot have phase separations on a shorter length scale if 
we are to stay in the `lowest band' of the lattice.
Let us now increase the interaction strengths such that 
`higher bands' must be taken into account, see, e.g., \cite{OL_us},  
to properly describe the ground state of the bosonic mixture.
We consider the filling factors $f_{A,B}=1$ and the intra-species parameters $g_A=g_B=10$.
Obviously, for $g_{AB}=0$ we have two strongly-interacting Mott-insulators which sit one atop the other.
This is also the situation for $g_{AB}=1$, see Fig.~3A,
and there is nothing in the {\it appearance} of these orbitals
to tell us what is about to happen when $g_{AB}$ is increased.
For $g_{AB}=10$ we do obtain phase separation on the {\it shortest} length scale,
namely inside a single lattice site, see Fig.~3B.
In each site, the two orbitals are pressed one against the other and thus narrow,
and localize at the borders of the lattice site.
In other words, the inter-species repulsion is sufficient to 
make the $A$ and $B$ bosons to separate {\it in} each site and
form hetero-atom 'pairs'.
What happens when $g_{AB}$ is further increased?
For $g_{AB}=30$ we get an alternating arrangement of
now homo-atom 'pairs', see Fig.~3C.
Finally, for $g_{AB}=40$ the two strongly-interacting
Mott-insulators fully separate in space as shown in Fig.~3D, 
which concludes our studies.

Let us briefly summarize. 
We have shown that bosonic mixtures in optical lattices
can undergo a wealth of phase separations on different length scales.
The stronger the intra-species interactions are the shorter is this length scale.
Specifically, (i) superfluids can only fully separate, i.e., occupy different regions of space; 
(ii) strongly-interacting superfluids can form alternating `domains' whose sizes
are a few lattice sites;
(iii) Mott-insulators can undergo demixing where the atoms of each species reside in neighboring lattice sites;
and, finally, (iv) strongly-interacting Mott-insulators can undergo demixing on the shortest length scale, 
namely, inside each lattice site.

\begin{figure}[ht]
\includegraphics[width=11cm,angle=-90]{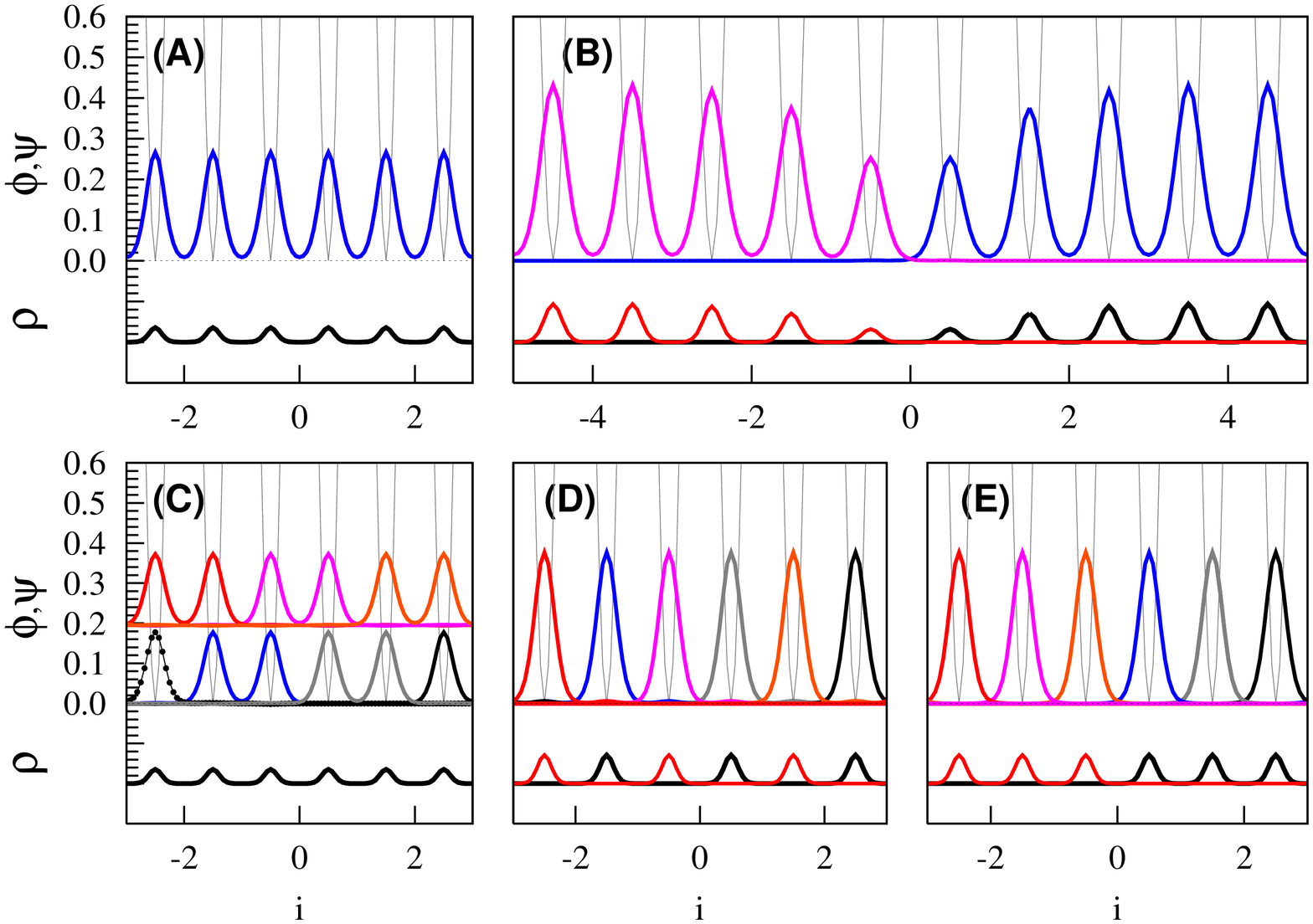}
\caption [kdv]{(Color online) Phase separation scenarios of two superfluids
for different intra- and inter-species interaction strengths.
Shown are the orbitals $\phi$ and $\psi$ of the $A$ and $B$ species, respectively,
and densities $\rho$ (lower curves).
Orbitals of $A$ bosons are in black, gray, and blue,
and of $B$ bosons -- in orange, magenta and red.
Densities are in black ($A$ bosons) and red ($B$ bosons).
The optical lattice is illustrated for guidance by the background sinusoidal curve.
The index ``i'' enumerates lattice maxima.
For convenience, the $\sqrt{n_i/N_A}\phi_i(x)$ 
and $\rho_A(x)= \frac{1}{N_A} \sum_i^{n_{orb}} n_i |\phi_i(x)|^2$ 
(scaled by $\frac{1}{2}$) are plotted, 
and similarly for the $B$ bosons.
All figures are for the filling factors $f_A=f_B=\frac{1}{2}$
in an optical lattice of $E_R=20$ recoil energies in depth and $16$ lattice sites in length.
The intra- and inter-species coupling parameters are:
(A) $g_A=g_B=10^{-3}$, $g_{AB}=10^{-4}$. Orbitals of both species sit one on top of the
other and cannot be distinguished;
(B) $g_A=g_B=10^{-3}$, $g_{AB}=1$;
(C) $g_A=g_B=0.1$, $g_{AB}=10^{-3}$. Orbitals of both species sit one on top of the
other with an 'offset' of one lattice site. For clarity of presentation
$B$ species orbitals are shifted upward and all orbitals are scaled by $\frac{2}{3}$;
(D) $g_A=g_B=0.1$, $g_{AB}=0.1$;
(E) $g_A=g_B=0.1$, $g_{AB}=10$.
}
\end{figure}

\begin{figure}[ht]
\includegraphics[width=11cm,angle=-90]{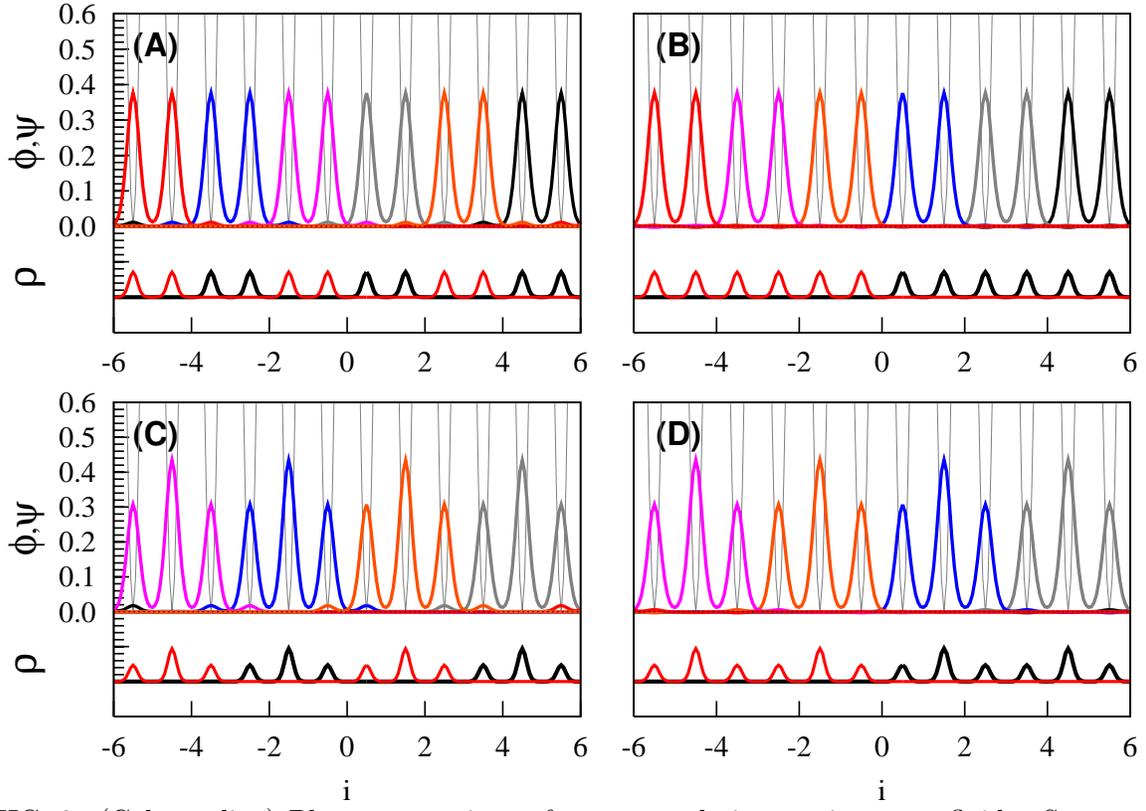}
\caption [kdv] {(Color online) Phase separations of two strongly-interacting superfluids.
Same as in Fig.~1 except for the parameters:
$g_A=g_B=0.1$ (all figures), 
and (A) $g_{AB}=0.1$, $f_A=f_B=\frac{1}{4}$;
(B) $g_{AB}=10$, $f_A=f_B=\frac{1}{4}$;
(C) $g_{AB}=0.1$, $f_A=f_B=\frac{1}{6}$ ($18$ lattice sites);
(D) $g_{AB}=10$, $f_A=f_B=\frac{1}{6}$ ($18$ lattice sites).
}
\end{figure}

\begin{figure}[ht]
\includegraphics[width=11cm,angle=-90]{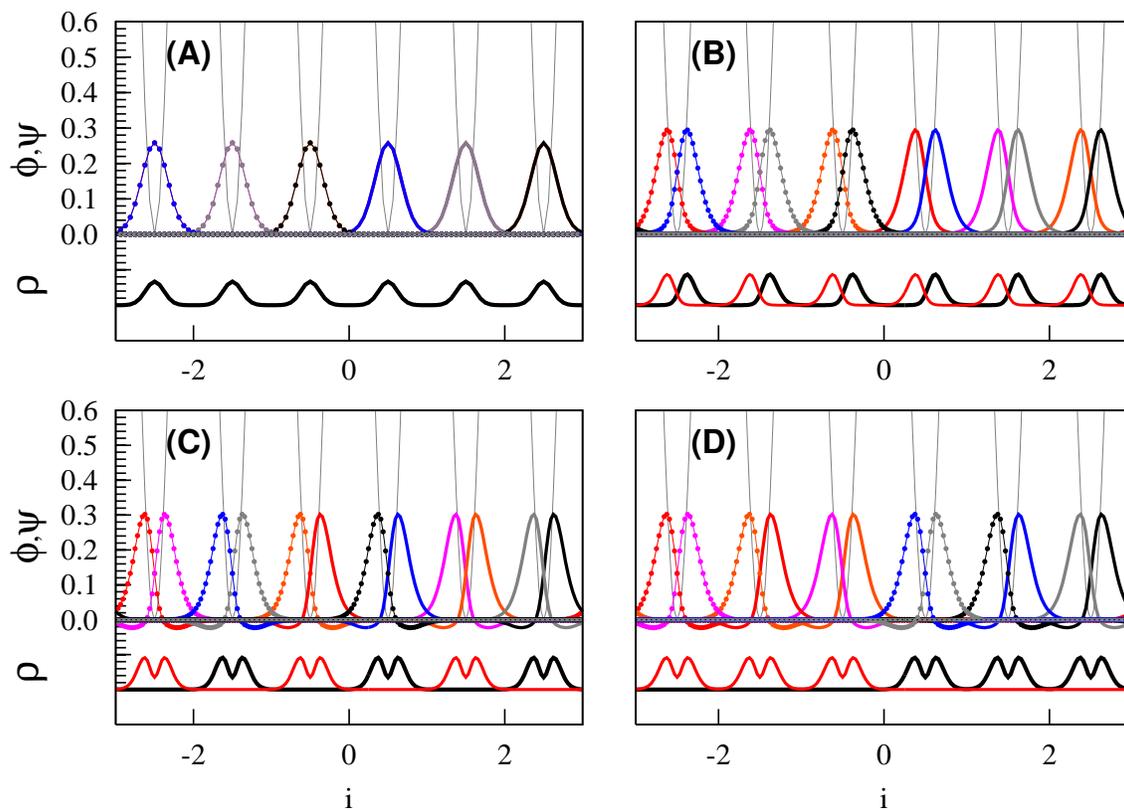}
\caption [kdv] {(Color online) Phase separations of two strongly-interacting Mott-insulators.
Same as in Fig.~1 except for the parameters:
$g_A=g_B=10$, $f_A=f_B=1$ and no scaling of $\rho$ (all figures), 
and (A) $g_{AB}=1$;
(B) $g_{AB}=10$;
(C) $g_{AB}=30$;
(D) $g_{AB}=40$.
}
\end{figure}

\ed